\documentclass[twocolumn,showpacs,preprintnumbers,amsmath,prl,amssymb,superscriptaddress]{revtex4-1}

\usepackage[export]{adjustbox}
\usepackage{amsfonts}
\usepackage{amsmath}
\usepackage[makeroom]{cancel}
\usepackage[english]{babel}
\usepackage[T1]{fontenc}
\usepackage{times}
\usepackage{mathrsfs}
\usepackage{graphicx}
\usepackage{dcolumn}
\usepackage{bm}
\usepackage{wasysym}
\usepackage[colorlinks,bookmarks=true,citecolor=blue,linkcolor=red,urlcolor=blue]{hyperref}
\usepackage[tight, FIGTOPCAP, hang, raggedright, nooneline]{subfigure}
\usepackage{hyperref}
\usepackage{xcolor}
\usepackage{epsfig}
\usepackage{amssymb}
\usepackage{lipsum}
\usepackage{dsfont}
\usepackage{color}
\usepackage{appendix}
\usepackage{epsfig}
\usepackage{tikz}
\usetikzlibrary{shapes.geometric}
\usetikzlibrary{decorations.markings}
\usetikzlibrary{decorations.pathmorphing}
\usepgflibrary{decorations.shapes}
\usepgflibrary{shapes.misc}

\begin{document}
\title{Quantized gravitational responses and the sign problem} 

\author{Z.~Ringel}
\affiliation{Rudolf Peierls Centre for Theoretical Physics, Keble Road, Oxford, OX1 3NP, United Kingdom}
\author{D.~L.~Kovrizhin}
\affiliation{Rudolf Peierls Centre for Theoretical Physics, Keble Road, Oxford, OX1 3NP, United Kingdom}
\affiliation{NRC Kurchatov institute, 1 Kurchatov sq., 123182, Moscow, Russia}

%
%


\maketitle

\textbf{
It is believed that not all quantum systems can be simulated efficiently using classical computational resources. This notion is supported by the fact that in quantum Monte Carlo (QMC) simulations for a large number of important problems it is not known how to express the partition function in a sign-free manner. The answer to the question --- whether there is an fundamental obstruction to such a sign-free representation in generic quantum systems --- remains unclear. Here, focussing on systems with bosonic degrees of freedom, we show that quantized gravitational responses appear as obstructions to local sign-free QMC. In condensed matter physics settings these responses, such as thermal Hall conductance, are associated with fractional quantum Hall effects. We show that similar arguments hold also in the case of spontaneously broken time-reversal (TR) symmetry such as in the chiral phase of a perturbed quantum Kagome antiferromagnet. The connection between quantized gravitational responses and the sign problem is also clearly manifested in certain vertex models, where TR symmetry is preserved.
}

There is a common wisdom related to computational complexity classes, that quantum mechanics is vastly superior to classical mechanics. Indeed all problems which can be solved efficiently on a classical computer can be solved as efficiently on a quantum computer, the opposite is believed not to hold. From condensed matter theory perspective, natural computational problems arise in the studies of models of interacting quantum many-body systems. Notably, some of these systems with purely bosonic degrees of freedom can be used for universal quantum computations \cite{Joost2001, Kolodrubetz2014}, -- therefore simulating them classically in polynomial time is likely to be impossible. Indeed, despite decades of efforts, efficient classical simulators are not known for many physically relevant bosonic models. Among those are the antiferromagnetic Heisenberg model on a Kagome lattice, and most, but not all, see e.g.~\cite{Mont2013, Gazit2016}, bosonic fractional quantum Hall effects (FQHE). This state of affairs might be simply due to a lack of one's analytical ingenuity. However one cannot exclude a possibility that some inherent physical property creates an obstruction to efficient classical simulations of many-body quantum systems.  

Establishing an obstruction to a classical simulation is a rather ill-defined task. A related, yet more concrete, goal is to find an obstruction to an efficient QMC, which is one of the chief numerical workhorses in the field. Specifically, one considers the Euclidean (or thermal) partition function of a quantum system in $d$-dimensions as a statistical mechanical (SM) model in $d+1$ dimensions with the extra dimension being imaginary time \cite{Cardy}. If this can be done such that the resulting SM model has Boltzmann weights which are local, and non-negative, then efficient QMC sampling is possible \cite{fnote1,Troyer:2005yq}. 

\begin{figure*}[tb!]
\centering
\includegraphics[width=1.5\columnwidth]{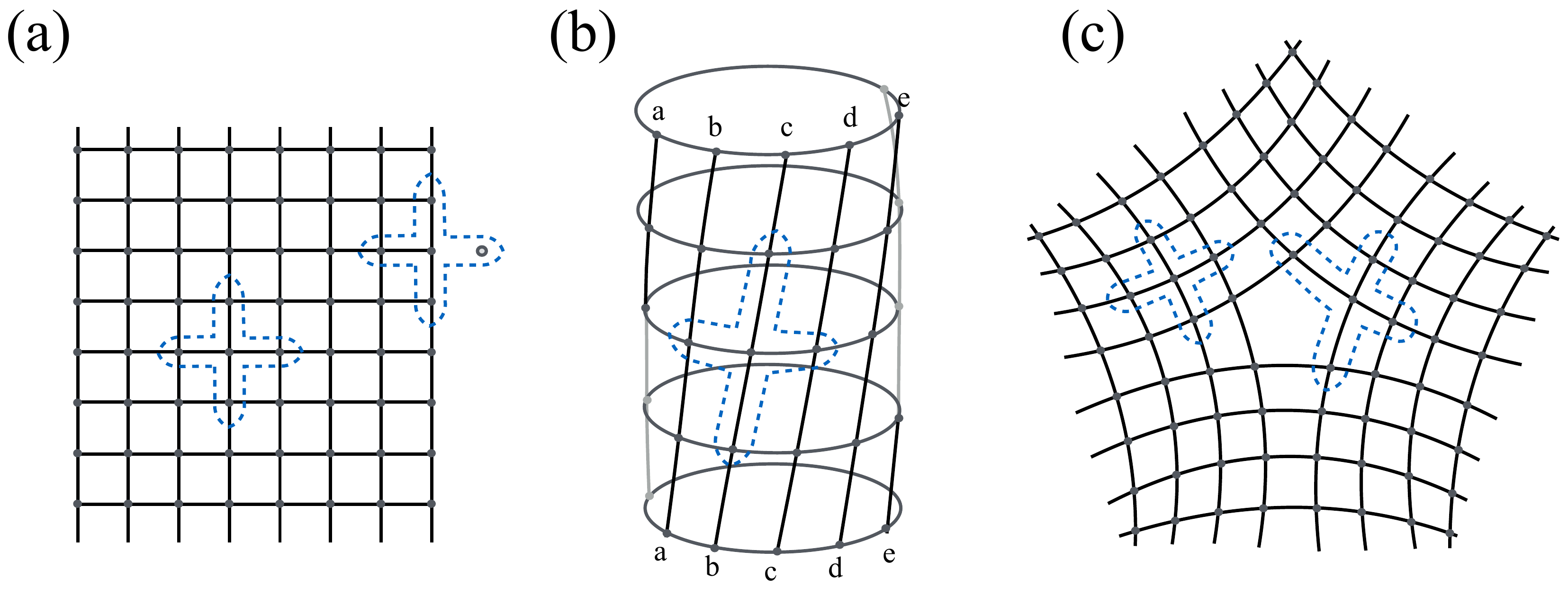}
\caption{{\bf Sign free manipulations of 2D lattices.} All figures show lattices with on-site bosonic degrees of freedom (full circles) with non-negative Boltzmann weights (dashed lines) involving several neighbouring sites. (a) open boundaries can be introduced in a sign-free manner by adding fictitious ghost degrees of freedom (empty circles). (b) global torsional twists are generated by reconnecting the lattice in a skewed manner. (c) local curvature is generated by disclinations. The latter can be introduced by lattice reconnections. Notwithstanding, in the fractional quantum Hall effect, a global torsional twist similar to (b) induces complex phases in the partition function.}\label{fig: twists}
\end{figure*}

While QMC offers essentially numerically-exact solution for a large variety of many-body quantum problems, in the case of aforementioned systems, the Boltzmann weights are in general negative or complex, and one has a ``sign problem''. We have to stress, see also Ref.~\cite{Hastings2015}, that the definition of a sign problem must involve the notion of locality. Indeed, by performing a non-local transformation on the physical degrees of freedom, one can always diagonalize the Hamiltonian. However, such a transformation requires computational resources which scale exponentially with the system size. Therefore for a meaningful definition of the sign problem one has to add an additional requirement that the degrees of freedom $\sigma$, which enter the Boltzmann weights, must be expressed as \textit{local} combinations of the physical ones, or more precisely that $\sigma$ are given by finite depth local quantum circuits acting on the physical degrees of freedom. Note that in this work we do not address a possibility of non-local approaches to solutions of the sign problem, such as determinant QMC, or cluster algorithms \cite{Chandrasekharan:1999rt,Cox:2000fr,Nyfeler:2008fb}.

In a recent paper, Hastings showed that for Hamiltonians consisting of commuting projectors, in the doubled semion phase, the partition function has a sign problem. Though not strictly proven, it is likely that this sign problem is the property of the doubled semion phase (or more precisely its $S-$matrix), that holds regardless of the requirement of having commuting projectors. Another interesting insight comes from a recent work by Troyer et.al.~\cite{Troyer}, who argued that the sign problem in auxillary-field determinant QMC arises from Aharonov-Bohm like phases, and thus has a topological origin. In our work, building on some of these insights, we show that in bosonic systems with quantized gravitational responses the sign problem cannot be cured by local transformations of degrees of freedom. We note that we study bosonic systems here, as opposed to fermions, in order to highlight that the sign problem which we discuss here is not related to the trivial fermionic exchange.

We start with the case of fractional quantum Hall effect. A hint to what property of FQHE states is in tension with a sign-free QMC comes from recent studies of bosonic symmetry protected topological states (SPT) \cite{Mont2013,Scaffidi2016,Bondesan2016, Gazit2016}. The latter found that certain bilayer bosonic integer quantum Hall effects (characterised by a non-zero Hall conductance) allow one for a sign-free QMC implemented via a local change of basis \cite{Scaffidi2016}. Since both, the FQHE and these SPTs, show quantized bulk electromagnetic responses \cite{Ryu2016}, it is not obvious how to draw an electromagnetic distinction between them. However, a distinction between these phases does exist, and is highlighted by the responses to temperature gradients, or via quantized gravitational responses \cite{Ryu2016}. Unlike the SPTs, FQHEs show a non-zero thermal Hall conductance. 

Interestingly, another canonical example of a sign problem arises in the studies of quantum Kagome lattice antiferromagnets (K-AFM). Some of these models have close connections to the physics of FQHE. For example, it was recently found that the ground state of the perturbed K-AFM model \cite{Wietek:2015fj,Trebst2014}, lies in proximity of the Kalmeyer-Laughlin state \cite{He2014,Kalmeyer1987}, which is known to possess non-trivial gravitational responses. 

Turning to a different set of models -- critical vertex models representing unitary minimal conformal field theories -- gravitational responses yet again appear to be indicative of a sign problem. While unitary minimal models have a sign-free formulation, no such formulations are known for their vertex counterparts with larger operator content \cite{FODA1989}. Interestingly, the low energy theories of these vertex models contain a term which attaches electric charge to the curvature \cite{DiFran,Kostov2000} -- reminiscent of the Wen-Zee term in the FQHE \cite{Wen1992}, whose coefficient is proportional to the non-chiral thermal Hall conductance.

\textit{Outline.} In this article we discuss the tension between gravitational responses and the sign problem in QMC. We show that \textit{no local sign-free QMC formulation is possible for phases having non-trivial gravitational responses}. 
Before delving into a detailed proof, let us sketch the main ideas. Just like Hall conductance, the thermal Hall conductance is associated with a quantum anomaly -- a gravitational anomaly. Generically, when coupling anomalous theories to static gauge fields, one finds that fluxes of the gauge field induce quantized complex phases in the partition function \cite{Ryu2016}. However, these complex phases do not immediately imply a sign problem, as the latter may not come from the pristine theory, but rather from its coupling to the gauge fields. Turning to gravity one can view the diffeomorphism invariance as a gauge symmetry \cite{Nakahara2003}, whose fluxes are the torsion, and the curvature. Conveniently for our purposes, the coupling to the torsion, and the curvature can always be introduced via lattice defects without generating signs or complex phases, see Fig.~\ref{fig: twists}. This observation suggests the link between the presence of gravitational anomaly, and the sign problem in the pristine theory. 

Let us make more precise our setting. In general, a Euclidean partition function of a $d$-dimensional quantum problem can be written as a $d+1$-dimensional partition function of a corresponding classical problem on a periodic $d+1$-dimensional lattice with degrees of freedom $\sigma$ attached to the sites of the lattice. We will call a partition function \textit{classical} if it is given by a product of Boltzmann weights, being real non-negative functions of $\sigma$'s. These weights must be local, and are functions of $\sigma$'s which involve a finite number of nearby sites. Further, it is useful to assume that a discrete translational invariance holds with some finite, lattice vector.

We start with an observation that a classical partition function on the plane can be transported to the torus, or to a twisted torus, as for both the lattice appears locally everywhere as a plane, see Fig.~\ref{fig: twists}. Further, the partition function can be used to construct a non-negative transfer operator ${\rm T}$ along the periodic (imaginary time) direction, such that the partition function is given by ${\rm T}$ raised to a suitable power. We shall always pick the transfer-matrix direction along the imaginary time axis.  
 
Let us assume that a classical 2+1 dimensional Euclidean partition function exists for the bosonic FQHE  (e.g. at $\nu=1/2$) on a plane or a torus. If this is the case the partition function can be transported to the following geometry -- a time-space $3D$ bulk on a lattice, open boundary conditions in the $y$-direction, and periodic boundary conditions with period $L$, and $\beta$ in the $x$, and the imaginary time $z$ directions respectively (we assume a large $\beta$, i.e.~larger then any other parameter.). We further introduce a twist in the boundary conditions such that $(x+\phi L,y,z) \equiv (x,y,z+\beta)$. Following \cite{Ryu2016} the resulting partition function on the left/right edge transforms under an insertion of a modular/torsional transformation ($\phi \rightarrow \phi + \alpha$) in the $xz-$plane as 
$Z_{l,r}[\phi+\alpha] = e^{\pm i\pi\alpha/12}Z_{l,r}[\phi]$.
The total partition function being the product of the contributions from the left and the right edge  has canceling phase factors.  The equation above is understood as a gravitational anomaly occurring within each edge \cite{Ryu2016}. This is analogous to case of the $U(1)$ charge anomaly \cite{Ryu2016,Ringel2013}, where the charge, despite being conserved on each edge separately, gets redistributed between two edges after a flux insertion. %

The tension between sign-free partition function, and the above phase factor is sharpened by the following observations. First, we recall that if the partition function can be made positive for $\phi=0$ (regular torus), it can also be made positive for $\phi \neq 0$ (twisted torus), and in particular the transfer operator remains real and non-negative everywhere, see Fig.~\ref{fig: twists}. This implies that one cannot have a sign-free partition function describing a single edge. Still in the full partition function this tension is obscured by the canceling phase factors arising from both edges. However, in the presence of a gapped bulk we can unambiguously separate the low energy contributions from left/right edges. We show below that while the phase contributions in the total partition function cancel, they nonetheless imply that the transfer operator must contain non-trivial phase factors. 

Let us outline two ways of formally achieving this left/right separation. The first, which makes a direct connection with the gravitational anomaly, requires augmenting the twist such that it acts only on a single edge, see Appendix II. The second more physical way is to note that pure gravitational anomalies are always associated with chiral systems. It is thus sufficient to prove that a sign-free formulation is in conflict with having only chiral excitations on a single edge. 

\textit{Sketch of proof.} We will use a proof by contradiction. Consider a transfer operator $\mathrm{T}$ for a FQHE Hamiltonian on a lattice, and \textit{assume that $\mathrm{T}$ is real, and non-negative}. Since $\mathrm{T}$ commutes with $e^{i a P}$, where $P$ is the lattice momentum operator (if a sublattice structure is present we interpret $a$ and $P$ accordingly), we can classify eigenstates of the transfer operator using momentum eigenvalues, and because $\mathrm{T}$ is given by a real matrix; the eigenvalues,  momenta, and eigenstates must come in complex-conjugate pairs $\{\lambda_n,e^{ik_n a}, |n\rangle\}$,$\{\lambda^*_n,e^{-ik_n a},|\bar{n} \rangle\}$. Notably if $k_n=0$ it is possible for the state to be its own complex conjugate. The non-negativity of $\mathrm{T}$ together with the known fact that FQHE ground states on a cylinder are unique (non-degenerate), implies that the largest eigenvalue $\lambda_0$ is positive and therefore gives real minimum ground state energy $\varepsilon_0$. The non-negativity of the ground state energy, and of the operator $e^{i a P}$ further implies that $e^{ik_0 a}=1$.

Next we discuss excited states, and their identification in the eigenvalues of the transfer operator. On general grounds one expects the highest weight states ($\lambda_n$) of ${\rm T}$ to be in one-to-one correspondence with the low energy spectrum of the Hamiltonian \cite{Faddeev1996,Kim1987} based on $\varepsilon_n = -\epsilon^{-1}{\mathrm{Re}}[\ln(\lambda_n)]$, where $\epsilon$ denotes a lattice spacing in the time-direction. Since we are mainly interested in the simulation of quantum Hamiltonians, transfer matrices would appear as $e^{-\epsilon H}$, or more generally $\Pi_i e^{-\epsilon H_i}$ where $H = \sum_i H_i$, and $\epsilon$ is small enough to ensure that the low energy physics is faithfully reproduced. More formally we thus limit ourselves to the possibility of solving a sign-problem by performing a local similarity transformation $S$, so that the transfer matrix has  the form $S \Pi_i e^{-\epsilon H_i} S^{-1}$. It is clear in this representation that the low energy spectra of $H$, and of ${\rm T}$ coincide up to terms of the order $O(\epsilon^2)$. 

Following this correspondence the chiral nature of FQHE edge states implies that there are two lowest collective excitations $|\varepsilon_q, e^{iqa},l\rangle,|\varepsilon_n, e^{-iqa},r\rangle$, which are characterised by the eigenvalues of $e^{iaP}$, i.e.~$e^{+i q a}$ on the left edge, and $e^{-i q a}$ on the right edge with $q=2\pi/L$, and corresponding energies $\varepsilon_{\pm q} = \varepsilon_0 + O(1/L)$. Let us be more precise on how edge labels are assigned. Given a collective excitation characterised by the eigenvector $|a\rangle$ of the transfer matrix, we compare the density matrix associated with the excitation $|a\rangle$, on the left or right edge $\rho_{l/r}[|a\rangle]$, with the reduced density matrix of the ground state $\rho_{l/r}[|gs\rangle]$. If, for example, $\rho_{l}[|a\rangle] =\rho_{l}[|gs\rangle]$ (up to factors $\sim e^{-L/l_B}$, where $l_B$ is the magnetic length) we say that the excitation $|a\rangle$ is localized on the right edge. Notably $\rho_{l/r}[|a\rangle^*] = \rho_{l/r}[|a \rangle]^*$, and since $|gs\rangle = |gs \rangle^*$ we find that $\rho_{l/r}[|gs \rangle]$ is real. Therefore, complex conjugation of the eigenstate of the low-energy collective excitation associated with one edge cannot be transformed into an excitation on the opposite edge. 

Finally, we note that given an excitation $|\varepsilon_{q},e^{iqa},l\rangle$, an orthogonal state with a complex conjugate momentum $e^{-iqa}$ must appear on the \textit{same} edge. However given chiral nature of edge excitations, we know that there is no such state in the spectrum and we thus arrived at the desired contradiction, which completes our proof. %

\textit{Spontaneously broken TR symmetry.} Having considered FQHE states, which are induced by an external magnetic field, we turn to discuss cases where FQHE-like states can appear as a result of spontaneous symmetry breaking. We will discuss a possibility of such a situation in the case of frustrated Heisenberg models, such as the Kagome quantum antiferromagnets. The latter are known for having a sign problem. Interestingly frustration may give rise to exotic spin-liquid states at low temperatures including chiral spin-liquids \cite{Wietek:2015fj,Trebst2014}. One of the candidate for these states is described by the Kalmeyer-Laughlin wave-function \cite{Kalmeyer1987}. The latter is in the same universality class as the bosonic $\nu=1/2$ FQHE. There is a numerical evidence that a perturbed Kagome quantum antiferromagnet (K-AFM') in presence of next, and next-next nearest neighbour Ising interactions \cite{He2014} is described by Kalmeyer-Laughlin phase.

Let us discuss a spontaneous TR symmetry breaking scenario on the example of K-AFM' model. A standard prescription for observing a broken symmetry is to couple the system to an infinitesimal ordering field. However, we want to avoid adding such operators here because they can introduce sign-problems. Therefore we use Anderson tower of states picture as an alternative indicator of a possibility for having broken symmetry. Now, because the model cannot strictly speaking break the time-reversal symmetry, its low lying spectrum cannot be chiral. Indeed, each of the two symmetry-broken lowest energy states $|gs\rangle$, and $|\bar{gs}\rangle$ support either chiral or anti-chiral edge excitations. Consequently, an excited state produced by complex conjugating a chiral edge excitation (belonging to one of the symmetry-broken states) \textit{does} appear in the spectrum on the same edge (c.f. with the FQHE case), but must be associated with the other symmetry-broken state. Thus the argument presented above for FQHE does not straightforwardly apply in this case.

Let us first assume that $|gs\rangle$ is proportional to $K|\bar{gs}\rangle$ on the computational basis. We claim that this cannot occur if $\mathrm{T}$ is non-negative, regardless of any chiral or topological physics.  To give an intuitive example consider a simple $1D$ Ising ferromagnet. On the $S_z$ basis $\mathrm{T}$ is non-negative. However, the two ground states are not connected by complex conjugation. On the other hand, on the $S_y$ basis the states are connected by $K$, but at the price of having negative entries in $\mathrm{T}$. We defer the proof for a more general case to the Appendix I. 

In the complementary case of $|gs\rangle$ being orthogonal to $K|\bar{gs}\rangle$ it is natural to assume that these two states will be distinct in the bulk, and our argument for FQHE holds.

\textit{Vertex models.}
Finally, we address the sign problem appearing in certain vertex models. Let us consider the six-vertex model formulation of the $2+0D$ minimal conformal field theory \cite{Zinn2009}. The six-vertex model can be thought of as a dense loop model, for example, by splitting every vertex into two loop configurations. It is known that for positive integer loop fugacities $n$ one can write a local partition function with positive Boltzmann weights \cite{Zinn2009}. However, in the general case one has to resort to complex Boltzmann weights.  

The Ising critical point can be described by an SM model, therefore it does not have a sign problem. Notwithstanding, to the best of our knowledge, no local sign-free reformulation is known for its loop model counterpart, for which $n=\sqrt{2}$. To clarify this, we note that while the equality between the Ising model and the vertex model holds at the level of partition functions, their operator content is different, and the equality of their partition functions is a result of miraculous cancellation of a large number of contributions in the vertex model case, see Ref.~\cite{FODA1989}. 

Similarly the field theories governing the two models are distinct. The Ising model is described by a scalar field $m$, while the vertex model is described by a compact boson $\phi$ coupled to gaussian curvature \cite{DiFran}. Thus a curvature defect, or a disclination on a lattice \cite{Kostov2000}, binds an electric charge in the vertex model, but no similar effect is expected for the Ising model. Thus in this very different setting a sign problem comes accompanied by a gravitational response. It is worth mentioning that: (i) local operators in the vertex model are vortices in $\phi$ (the magnetic-charges), therefore disclinations, by binding electric charges, couple to them via Aharonov-Bohm fluxes, and (ii) disclinations can be added without introducing signs into partition functions, see Fig.~\ref{fig: twists}.

\textit{Discussion.} To summarize we suggest that non-trivial gravitational responses can be identified with obstructions to  sign-free local quantum Monte Carlo implementations. We substantiate this idea using examples of fractional quantum Hall effects, frustrated quantum magnets, and vertex models. Curiously, Tensor network based numerical approaches, for which the sign problem is irrelevant by construction, also struggle with simulating FQHE states (in the thermodynamic limit). This raises an intriguing connection between gravity and computational complexity via sign-problems~\cite{Troyer:2005yq}.
 
\textit{Acknowledgements.} We would like to thank J.T.~Chalker and Andr\'e Lukas for helpful discussions. We also like to thank A.~L\"auchli, and S.~Gazit for useful suggestions and comments on the manuscript. Z.~R.~was supported by the European Union's Horizon 2020 research and innovation programme under the Marie Sklodowska-Curie grant agreement No.~657111. D.K.~was supported by the EPSRC Grant No.~EP/M007928/1.

\bibliography{Draft}

\begin{thebibliography}{28}%
\makeatletter
\providecommand \@ifxundefined [1]{%
 \@ifx{#1\undefined}
}%
\providecommand \@ifnum [1]{%
 \ifnum #1\expandafter \@firstoftwo
 \else \expandafter \@secondoftwo
 \fi
}%
\providecommand \@ifx [1]{%
 \ifx #1\expandafter \@firstoftwo
 \else \expandafter \@secondoftwo
 \fi
}%
\providecommand \natexlab [1]{#1}%
\providecommand \enquote  [1]{``#1''}%
\providecommand \bibnamefont  [1]{#1}%
\providecommand \bibfnamefont [1]{#1}%
\providecommand \citenamefont [1]{#1}%
\providecommand \href@noop [0]{\@secondoftwo}%
\providecommand \href [0]{\begingroup \@sanitize@url \@href}%
\providecommand \@href[1]{\@@startlink{#1}\@@href}%
\providecommand \@@href[1]{\endgroup#1\@@endlink}%
\providecommand \@sanitize@url [0]{\catcode `\\12\catcode `\$12\catcode
  `\&12\catcode `\#12\catcode `\^12\catcode `\_12\catcode `\%12\relax}%
\providecommand \@@startlink[1]{}%
\providecommand \@@endlink[0]{}%
\providecommand \url  [0]{\begingroup\@sanitize@url \@url }%
\providecommand \@url [1]{\endgroup\@href {#1}{\urlprefix }}%
\providecommand \urlprefix  [0]{URL }%
\providecommand \Eprint [0]{\href }%
\providecommand \doibase [0]{http://dx.doi.org/}%
\providecommand \selectlanguage [0]{\@gobble}%
\providecommand \bibinfo  [0]{\@secondoftwo}%
\providecommand \bibfield  [0]{\@secondoftwo}%
\providecommand \translation [1]{[#1]}%
\providecommand \BibitemOpen [0]{}%
\providecommand \bibitemStop [0]{}%
\providecommand \bibitemNoStop [0]{.\EOS\space}%
\providecommand \EOS [0]{\spacefactor3000\relax}%
\providecommand \BibitemShut  [1]{\csname bibitem#1\endcsname}%
\let\auto@bib@innerbib\@empty
\bibitem [{\citenamefont {{Slingerland}}\ and\ \citenamefont
  {{Bais}}(2001)}]{Joost2001}%
  \BibitemOpen
  \bibfield  {author} {\bibinfo {author} {\bibfnamefont {J.~K.}\ \bibnamefont
  {{Slingerland}}}\ and\ \bibinfo {author} {\bibfnamefont {F.~A.}\ \bibnamefont
  {{Bais}}},\ }\href {\doibase 10.1016/S0550-3213(01)00308-X} {\bibfield
  {journal} {\bibinfo  {journal} {Nuclear Physics B}\ }\textbf {\bibinfo
  {volume} {612}},\ \bibinfo {pages} {229} (\bibinfo {year} {2001})},\ \Eprint
  {http://arxiv.org/abs/cond-mat/0104035} {cond-mat/0104035} \BibitemShut
  {NoStop}%
\bibitem [{\citenamefont {Kolodrubetz}(2014)}]{Kolodrubetz2014}%
  \BibitemOpen
  \bibfield  {author} {\bibinfo {author} {\bibfnamefont {M.}~\bibnamefont
  {Kolodrubetz}},\ }\href {\doibase 10.1103/PhysRevB.89.045107} {\bibfield
  {journal} {\bibinfo  {journal} {Phys. Rev. B}\ }\textbf {\bibinfo {volume}
  {89}},\ \bibinfo {pages} {045107} (\bibinfo {year} {2014})}\BibitemShut
  {NoStop}%
\bibitem [{\citenamefont {Geraedts}\ and\ \citenamefont
  {Motrunich}(2013)}]{Mont2013}%
  \BibitemOpen
  \bibfield  {author} {\bibinfo {author} {\bibfnamefont {S.~D.}\ \bibnamefont
  {Geraedts}}\ and\ \bibinfo {author} {\bibfnamefont {O.~I.}\ \bibnamefont
  {Motrunich}},\ }\href {\doibase http://dx.doi.org/10.1016/j.aop.2013.03.017}
  {\bibfield  {journal} {\bibinfo  {journal} {Annals of Physics}\ }\textbf
  {\bibinfo {volume} {334}},\ \bibinfo {pages} {288 } (\bibinfo {year}
  {2013})}\BibitemShut {NoStop}%
\bibitem [{\citenamefont {Gazit}\ and\ \citenamefont
  {Vishwanath}(2016)}]{Gazit2016}%
  \BibitemOpen
  \bibfield  {author} {\bibinfo {author} {\bibfnamefont {S.}~\bibnamefont
  {Gazit}}\ and\ \bibinfo {author} {\bibfnamefont {A.}~\bibnamefont
  {Vishwanath}},\ }\href {\doibase 10.1103/PhysRevB.93.115146} {\bibfield
  {journal} {\bibinfo  {journal} {Phys. Rev. B}\ }\textbf {\bibinfo {volume}
  {93}},\ \bibinfo {pages} {115146} (\bibinfo {year} {2016})}\BibitemShut
  {NoStop}%
\bibitem [{\citenamefont {Cardy}(1996)}]{Cardy}%
  \BibitemOpen
  \bibfield  {author} {\bibinfo {author} {\bibfnamefont {J.}~\bibnamefont
  {Cardy}},\ }\href@noop {} {\emph {\bibinfo {title} {Scaling and
  Renormalization in Statistical Physics}}},\ Cambridge Lecture Notes in
  Physics\ (\bibinfo {year} {1996})\BibitemShut {NoStop}%
\bibitem [{Note1()}]{fnote1}%
  \BibitemOpen
  \bibinfo {note} {We note that glassiness may still present an
  obstruction}\BibitemShut {NoStop}%
\bibitem [{\citenamefont {Troyer}\ and\ \citenamefont
  {Wiese}(2005)}]{Troyer:2005yq}%
  \BibitemOpen
  \bibfield  {author} {\bibinfo {author} {\bibfnamefont {M.}~\bibnamefont
  {Troyer}}\ and\ \bibinfo {author} {\bibfnamefont {U.-J.}\ \bibnamefont
  {Wiese}},\ }\href {https://link.aps.org/doi/10.1103/PhysRevLett.94.170201}
  {\bibfield  {journal} {\bibinfo  {journal} {Physical Review Letters}\
  }\textbf {\bibinfo {volume} {94}},\ \bibinfo {pages} {170201} (\bibinfo
  {year} {2005})}\BibitemShut {NoStop}%
\bibitem [{\citenamefont {{Hastings}}(2016)}]{Hastings2015}%
  \BibitemOpen
  \bibfield  {author} {\bibinfo {author} {\bibfnamefont {M.~B.}\ \bibnamefont
  {{Hastings}}},\ }\href {\doibase 10.1063/1.4936216} {\bibfield  {journal}
  {\bibinfo  {journal} {Journal of Mathematical Physics}\ }\textbf {\bibinfo
  {volume} {57}},\ \bibinfo {eid} {015210} (\bibinfo {year} {2016})},\ \Eprint
  {http://arxiv.org/abs/1506.08883} {arXiv:1506.08883 [quant-ph]} \BibitemShut
  {NoStop}%
\bibitem [{\citenamefont {Chandrasekharan}\ and\ \citenamefont
  {Wiese}(1999)}]{Chandrasekharan:1999rt}%
  \BibitemOpen
  \bibfield  {author} {\bibinfo {author} {\bibfnamefont {S.}~\bibnamefont
  {Chandrasekharan}}\ and\ \bibinfo {author} {\bibfnamefont {U.-J.}\
  \bibnamefont {Wiese}},\ }\href
  {https://link.aps.org/doi/10.1103/PhysRevLett.83.3116} {\bibfield  {journal}
  {\bibinfo  {journal} {Physical Review Letters}\ }\textbf {\bibinfo {volume}
  {83}},\ \bibinfo {pages} {3116} (\bibinfo {year} {1999})}\BibitemShut
  {NoStop}%
\bibitem [{\citenamefont {Cox}\ \emph {et~al.}(2000)\citenamefont {Cox},
  \citenamefont {Gattringer}, \citenamefont {Holland}, \citenamefont
  {Scarlet},\ and\ \citenamefont {Wiese}}]{Cox:2000fr}%
  \BibitemOpen
  \bibfield  {author} {\bibinfo {author} {\bibfnamefont {J.}~\bibnamefont
  {Cox}}, \bibinfo {author} {\bibfnamefont {C.}~\bibnamefont {Gattringer}},
  \bibinfo {author} {\bibfnamefont {K.}~\bibnamefont {Holland}}, \bibinfo
  {author} {\bibfnamefont {B.}~\bibnamefont {Scarlet}}, \ and\ \bibinfo
  {author} {\bibfnamefont {U.~J.}\ \bibnamefont {Wiese}},\ }\bibfield
  {booktitle} {\emph {\bibinfo {booktitle} {Proceedings of the XVIIth
  International Symposium on Lattice Field Theory}},\ }\href {\doibase
  http://dx.doi.org/10.1016/S0920-5632(00)91804-8} {\bibfield  {journal}
  {\bibinfo  {journal} {Nuclear Physics B - Proceedings Supplements}\ }\textbf
  {\bibinfo {volume} {83}},\ \bibinfo {pages} {777} (\bibinfo {year}
  {2000})}\BibitemShut {NoStop}%
\bibitem [{\citenamefont {Nyfeler}\ \emph {et~al.}(2008)\citenamefont
  {Nyfeler}, \citenamefont {Jiang}, \citenamefont {K{\"a}mpfer},\ and\
  \citenamefont {Wiese}}]{Nyfeler:2008fb}%
  \BibitemOpen
  \bibfield  {author} {\bibinfo {author} {\bibfnamefont {M.}~\bibnamefont
  {Nyfeler}}, \bibinfo {author} {\bibfnamefont {F.~J.}\ \bibnamefont {Jiang}},
  \bibinfo {author} {\bibfnamefont {F.}~\bibnamefont {K{\"a}mpfer}}, \ and\
  \bibinfo {author} {\bibfnamefont {U.~J.}\ \bibnamefont {Wiese}},\ }\href
  {https://link.aps.org/doi/10.1103/PhysRevLett.100.247206} {\bibfield
  {journal} {\bibinfo  {journal} {Physical Review Letters}\ }\textbf {\bibinfo
  {volume} {100}},\ \bibinfo {pages} {247206} (\bibinfo {year}
  {2008})}\BibitemShut {NoStop}%
\bibitem [{\citenamefont {Iazzi}\ \emph {et~al.}(2016)\citenamefont {Iazzi},
  \citenamefont {Soluyanov},\ and\ \citenamefont {Troyer}}]{Troyer}%
  \BibitemOpen
  \bibfield  {author} {\bibinfo {author} {\bibfnamefont {M.}~\bibnamefont
  {Iazzi}}, \bibinfo {author} {\bibfnamefont {A.~A.}\ \bibnamefont
  {Soluyanov}}, \ and\ \bibinfo {author} {\bibfnamefont {M.}~\bibnamefont
  {Troyer}},\ }\href {\doibase 10.1103/PhysRevB.93.115102} {\bibfield
  {journal} {\bibinfo  {journal} {Phys. Rev. B}\ }\textbf {\bibinfo {volume}
  {93}},\ \bibinfo {pages} {115102} (\bibinfo {year} {2016})}\BibitemShut
  {NoStop}%
\bibitem [{\citenamefont {{Scaffidi}}\ and\ \citenamefont
  {{Ringel}}(2016)}]{Scaffidi2016}%
  \BibitemOpen
  \bibfield  {author} {\bibinfo {author} {\bibfnamefont {T.}~\bibnamefont
  {{Scaffidi}}}\ and\ \bibinfo {author} {\bibfnamefont {Z.}~\bibnamefont
  {{Ringel}}},\ }\href {\doibase 10.1103/PhysRevB.93.115105} {\bibfield
  {journal} {\bibinfo  {journal} {\prb}\ }\textbf {\bibinfo {volume} {93}},\
  \bibinfo {eid} {115105} (\bibinfo {year} {2016})},\ \Eprint
  {http://arxiv.org/abs/1505.02775} {arXiv:1505.02775 [cond-mat.str-el]}
  \BibitemShut {NoStop}%
\bibitem [{\citenamefont {{Bondesan}}\ and\ \citenamefont
  {{Ringel}}(2016)}]{Bondesan2016}%
  \BibitemOpen
  \bibfield  {author} {\bibinfo {author} {\bibfnamefont {R.}~\bibnamefont
  {{Bondesan}}}\ and\ \bibinfo {author} {\bibfnamefont {Z.}~\bibnamefont
  {{Ringel}}},\ }\href@noop {} {\bibfield  {journal} {\bibinfo  {journal}
  {ArXiv e-prints}\ } (\bibinfo {year} {2016})},\ \Eprint
  {http://arxiv.org/abs/1606.00447} {arXiv:1606.00447 [cond-mat.stat-mech]}
  \BibitemShut {NoStop}%
\bibitem [{\citenamefont {{Nakai}}\ \emph {et~al.}(2016)\citenamefont
  {{Nakai}}, \citenamefont {{Ryu}},\ and\ \citenamefont {{Nomura}}}]{Ryu2016}%
  \BibitemOpen
  \bibfield  {author} {\bibinfo {author} {\bibfnamefont {R.}~\bibnamefont
  {{Nakai}}}, \bibinfo {author} {\bibfnamefont {S.}~\bibnamefont {{Ryu}}}, \
  and\ \bibinfo {author} {\bibfnamefont {K.}~\bibnamefont {{Nomura}}},\
  }\href@noop {} {\bibfield  {journal} {\bibinfo  {journal} {ArXiv e-prints}\ }
  (\bibinfo {year} {2016})},\ \Eprint {http://arxiv.org/abs/1611.09463}
  {arXiv:1611.09463 [cond-mat.mes-hall]} \BibitemShut {NoStop}%
\bibitem [{\citenamefont {Wietek}\ \emph {et~al.}(2015)\citenamefont {Wietek},
  \citenamefont {Sterdyniak},\ and\ \citenamefont
  {L{\"a}uchli}}]{Wietek:2015fj}%
  \BibitemOpen
  \bibfield  {author} {\bibinfo {author} {\bibfnamefont {A.}~\bibnamefont
  {Wietek}}, \bibinfo {author} {\bibfnamefont {A.}~\bibnamefont {Sterdyniak}},
  \ and\ \bibinfo {author} {\bibfnamefont {A.~M.}\ \bibnamefont
  {L{\"a}uchli}},\ }\href {https://link.aps.org/doi/10.1103/PhysRevB.92.125122}
  {\bibfield  {journal} {\bibinfo  {journal} {Physical Review B}\ }\textbf
  {\bibinfo {volume} {92}},\ \bibinfo {pages} {125122} (\bibinfo {year}
  {2015})}\BibitemShut {NoStop}%
\bibitem [{\citenamefont {{Bauer}}\ \emph {et~al.}(2014)\citenamefont
  {{Bauer}}, \citenamefont {{Cincio}}, \citenamefont {{Keller}}, \citenamefont
  {{Dolfi}}, \citenamefont {{Vidal}}, \citenamefont {{Trebst}},\ and\
  \citenamefont {{Ludwig}}}]{Trebst2014}%
  \BibitemOpen
  \bibfield  {author} {\bibinfo {author} {\bibfnamefont {B.}~\bibnamefont
  {{Bauer}}}, \bibinfo {author} {\bibfnamefont {L.}~\bibnamefont {{Cincio}}},
  \bibinfo {author} {\bibfnamefont {B.~P.}\ \bibnamefont {{Keller}}}, \bibinfo
  {author} {\bibfnamefont {M.}~\bibnamefont {{Dolfi}}}, \bibinfo {author}
  {\bibfnamefont {G.}~\bibnamefont {{Vidal}}}, \bibinfo {author} {\bibfnamefont
  {S.}~\bibnamefont {{Trebst}}}, \ and\ \bibinfo {author} {\bibfnamefont
  {A.~W.~W.}\ \bibnamefont {{Ludwig}}},\ }\href {\doibase 10.1038/ncomms6137}
  {\bibfield  {journal} {\bibinfo  {journal} {Nature Communications}\ }\textbf
  {\bibinfo {volume} {5}},\ \bibinfo {eid} {5137} (\bibinfo {year} {2014})},\
  \Eprint {http://arxiv.org/abs/1401.3017} {arXiv:1401.3017 [cond-mat.str-el]}
  \BibitemShut {NoStop}%
\bibitem [{\citenamefont {{He}}\ \emph {et~al.}(2014)\citenamefont {{He}},
  \citenamefont {{Sheng}},\ and\ \citenamefont {{Chen}}}]{He2014}%
  \BibitemOpen
  \bibfield  {author} {\bibinfo {author} {\bibfnamefont {Y.-C.}\ \bibnamefont
  {{He}}}, \bibinfo {author} {\bibfnamefont {D.~N.}\ \bibnamefont {{Sheng}}}, \
  and\ \bibinfo {author} {\bibfnamefont {Y.}~\bibnamefont {{Chen}}},\ }\href
  {\doibase 10.1103/PhysRevLett.112.137202} {\bibfield  {journal} {\bibinfo
  {journal} {Physical Review Letters}\ }\textbf {\bibinfo {volume} {112}},\
  \bibinfo {eid} {137202} (\bibinfo {year} {2014})},\ \Eprint
  {http://arxiv.org/abs/1312.3461} {arXiv:1312.3461 [cond-mat.str-el]}
  \BibitemShut {NoStop}%
\bibitem [{\citenamefont {Kalmeyer}\ and\ \citenamefont
  {Laughlin}(1987)}]{Kalmeyer1987}%
  \BibitemOpen
  \bibfield  {author} {\bibinfo {author} {\bibfnamefont {V.}~\bibnamefont
  {Kalmeyer}}\ and\ \bibinfo {author} {\bibfnamefont {R.~B.}\ \bibnamefont
  {Laughlin}},\ }\href {\doibase 10.1103/PhysRevLett.59.2095} {\bibfield
  {journal} {\bibinfo  {journal} {Phys. Rev. Lett.}\ }\textbf {\bibinfo
  {volume} {59}},\ \bibinfo {pages} {2095} (\bibinfo {year}
  {1987})}\BibitemShut {NoStop}%
\bibitem [{\citenamefont {Foda}\ and\ \citenamefont
  {Nienhuis}(1989)}]{FODA1989}%
  \BibitemOpen
  \bibfield  {author} {\bibinfo {author} {\bibfnamefont {O.}~\bibnamefont
  {Foda}}\ and\ \bibinfo {author} {\bibfnamefont {B.}~\bibnamefont
  {Nienhuis}},\ }\href {\doibase
  http://dx.doi.org/10.1016/0550-3213(89)90525-7} {\bibfield  {journal}
  {\bibinfo  {journal} {Nuclear Physics B}\ }\textbf {\bibinfo {volume}
  {324}},\ \bibinfo {pages} {643 } (\bibinfo {year} {1989})}\BibitemShut
  {NoStop}%
\bibitem [{\citenamefont {Francesco}\ \emph {et~al.}(1997)\citenamefont
  {Francesco}, \citenamefont {Mathieu},\ and\ \citenamefont
  {Senechal}}]{DiFran}%
  \BibitemOpen
  \bibfield  {author} {\bibinfo {author} {\bibfnamefont {P.~D.}\ \bibnamefont
  {Francesco}}, \bibinfo {author} {\bibfnamefont {P.}~\bibnamefont {Mathieu}},
  \ and\ \bibinfo {author} {\bibfnamefont {D.}~\bibnamefont {Senechal}},\
  }\href {http://books.google.co.uk/books?id=keUrdME5rhIC} {\emph {\bibinfo
  {title} {Conformal Field Theory}}},\ Graduate Texts in Contemporary Physics\
  (\bibinfo  {publisher} {Springer},\ \bibinfo {year} {1997})\BibitemShut
  {NoStop}%
\bibitem [{\citenamefont {{Kostov}}(2000)}]{Kostov2000}%
  \BibitemOpen
  \bibfield  {author} {\bibinfo {author} {\bibfnamefont {I.~K.}\ \bibnamefont
  {{Kostov}}},\ }\href {\doibase 10.1016/S0550-3213(00)00060-2} {\bibfield
  {journal} {\bibinfo  {journal} {Nuclear Physics B}\ }\textbf {\bibinfo
  {volume} {575}},\ \bibinfo {pages} {513} (\bibinfo {year} {2000})},\ \Eprint
  {http://arxiv.org/abs/hep-th/9911023} {hep-th/9911023} \BibitemShut {NoStop}%
\bibitem [{\citenamefont {Wen}\ and\ \citenamefont {Zee}(1992)}]{Wen1992}%
  \BibitemOpen
  \bibfield  {author} {\bibinfo {author} {\bibfnamefont {X.~G.}\ \bibnamefont
  {Wen}}\ and\ \bibinfo {author} {\bibfnamefont {A.}~\bibnamefont {Zee}},\
  }\href {\doibase 10.1103/PhysRevLett.69.953} {\bibfield  {journal} {\bibinfo
  {journal} {Phys. Rev. Lett.}\ }\textbf {\bibinfo {volume} {69}},\ \bibinfo
  {pages} {953} (\bibinfo {year} {1992})}\BibitemShut {NoStop}%
\bibitem [{\citenamefont {Nakahara}(2003)}]{Nakahara2003}%
  \BibitemOpen
  \bibfield  {author} {\bibinfo {author} {\bibfnamefont {M.}~\bibnamefont
  {Nakahara}},\ }\href {https://books.google.co.uk/books?id=cH-XQB0Ex5wC}
  {\emph {\bibinfo {title} {Geometry, Topology and Physics, Second Edition}}},\
  Graduate student series in physics\ (\bibinfo  {publisher} {Taylor \&
  Francis},\ \bibinfo {year} {2003})\BibitemShut {NoStop}%
\bibitem [{\citenamefont {Ringel}\ and\ \citenamefont
  {Stern}(2013)}]{Ringel2013}%
  \BibitemOpen
  \bibfield  {author} {\bibinfo {author} {\bibfnamefont {Z.}~\bibnamefont
  {Ringel}}\ and\ \bibinfo {author} {\bibfnamefont {A.}~\bibnamefont {Stern}},\
  }\href {\doibase 10.1103/PhysRevB.88.115307} {\bibfield  {journal} {\bibinfo
  {journal} {Phys. Rev. B}\ }\textbf {\bibinfo {volume} {88}},\ \bibinfo
  {pages} {115307} (\bibinfo {year} {2013})}\BibitemShut {NoStop}%
\bibitem [{\citenamefont {{Faddeev}}(1996)}]{Faddeev1996}%
  \BibitemOpen
  \bibfield  {author} {\bibinfo {author} {\bibfnamefont {L.~D.}\ \bibnamefont
  {{Faddeev}}},\ }\href@noop {} {\bibfield  {journal} {\bibinfo  {journal}
  {ArXiv High Energy Physics - Theory e-prints}\ } (\bibinfo {year} {1996})},\
  \Eprint {http://arxiv.org/abs/hep-th/9605187} {hep-th/9605187} \BibitemShut
  {NoStop}%
\bibitem [{\citenamefont {Kim}\ and\ \citenamefont {Pearce}(1987)}]{Kim1987}%
  \BibitemOpen
  \bibfield  {author} {\bibinfo {author} {\bibfnamefont {D.}~\bibnamefont
  {Kim}}\ and\ \bibinfo {author} {\bibfnamefont {P.~A.}\ \bibnamefont
  {Pearce}},\ }\href {http://stacks.iop.org/0305-4470/20/i=7/a=006} {\bibfield
  {journal} {\bibinfo  {journal} {Journal of Physics A: Mathematical and
  General}\ }\textbf {\bibinfo {volume} {20}},\ \bibinfo {pages} {L451}
  (\bibinfo {year} {1987})}\BibitemShut {NoStop}%
\bibitem [{\citenamefont {{Zinn-Justin}}(2009)}]{Zinn2009}%
  \BibitemOpen
  \bibfield  {author} {\bibinfo {author} {\bibfnamefont {P.}~\bibnamefont
  {{Zinn-Justin}}},\ }\href@noop {} {\bibfield  {journal} {\bibinfo  {journal}
  {ArXiv e-prints}\ } (\bibinfo {year} {2009})},\ \Eprint
  {http://arxiv.org/abs/0901.0665} {arXiv:0901.0665 [math-ph]} \BibitemShut
  {NoStop}%
\end{thebibliography}%
\appendix

\section{``Complex conjugation symmetry'' in statistical mechanics}
A transfer matrix ${\mathrm T}$ of any SM model, being non-negative, possesses a global anti-unitary symmetry, which is defined by a complex conjugation operator $K$. It is a natural question to ask whether this symmetry can be broken spontaneously (see definitions below). We \textit{conjecture} that this is impossible in any physical setting. Specifically, we prove this conjecture in two relevant cases: (a) the model is inversion symmetric along some direction, such that ${\mathrm T}$ becomes a symmetric matrix, or (b) there is a degeneracy of the eigenvalues of the transfer matrix for broken symmetry ground states. We also discuss a complementary scenario where neither $(a)$ or $(b)$ hold.

Let us start by making a precise definition of what we mean by saying that complex conjugation symmetry is spontaneously broken: (i) the transfer matrices ${\mathrm T}(L)$, where $L$ is the system size, have two dominant eigenvalues $\lambda,\lambda'$ which are $O(e^{-L})$ apart in the logarithms of their magnitudes, and separated from all other states by a distance $\delta(L)$ which scales as $\delta(L)\sim L^{-a}$, where $a\ge0$; (ii) any linear superposition of $|\lambda\rangle,|\lambda'\rangle$ which is invariant under $K$ in the computational basis $\sigma$, is a Schroedinger cat state (cat state). By this we mean that it is an equal weight superposition of two distinct many-body states (physical states), or more precisely, the states which have non-zero matrix elements only for operators acting of $O(N)$ sites for a system of size $N$. (iii) The physical states $|\psi_1\rangle,|\psi_2\rangle$ map to each other under the action of $K$. (iv) The physical states are distinguishable by having finite, and different, expectation values for some local Hermitian operator $O_y$ which obeys $K O_y K = - O_y$ in the thermodynamic limit. 

First, we consider the simplest scenario where the absolute values of $\lambda,\lambda'$ are equal, $|\lambda|=|\lambda'|$. Then, by Perron-Frobenius theorem, this implies that $|\lambda \rangle$ and $|\lambda'\rangle$ may be chosen positive on the computational basis $|\sigma\rangle$ on which ${\rm T}$ is non-negative. We next show that having two non-negative ground state contradicts the above symmetry breaking assumptions. By these assumptions, we can write 
\begin{align}
|\lambda \rangle &= a |\psi_1\rangle +  b|\psi_2\rangle , \\ \nonumber
|\lambda' \rangle &= e^{i\varphi}(b^{*} |\psi_1\rangle - a^{*} |\psi_2\rangle) ,
\end{align}
where, $a,b$ are two complex numbers equal in magnitude, and $\varphi$ is an arbitrary phase. Note that both physical states, and the $\lambda$ states are assumed to be normalised. Now, using (ii) we have  
\begin{align}
\langle \lambda | O_y | \lambda' \rangle &=e^{i\phi}a b \left[ \langle\psi_1| O_y  | \psi_1\rangle - \langle \psi_2 | O_y  | \psi_2 \rangle\right]\\=& 2 e^{i\phi}a b\langle \psi_1 | O_y  | \psi_1 \rangle,
\end{align}
and from (iii)+(iv) we find that the r.h.s is non-zero. Next, let us split the computational basis $\sigma=(\eta,\mu)$, into the set of degrees of freedom on which $O_y$ acts like as an identity ($\eta$) and the set $\mu$ on which it acts non-trivially. We then find  
\begin{align}
|\langle \lambda | O_y | \lambda' \rangle| &= |\sum_{\eta,\mu,\mu'} \langle \lambda | \eta,\mu \rangle [O_y]_{\mu\mu'} \langle \eta,\mu'| \lambda' \rangle| \\ \nonumber &\leq \sum_{\eta,\mu,\mu'} \langle \lambda | \eta,\mu \rangle |[O_y]_{\mu\mu'}| \langle \eta,\mu'| \lambda' \rangle, 
\end{align}
where in the last line we have used the non-negativity of $|\lambda\rangle,|\lambda'\rangle$ on the $\sigma$ basis. Let us denote the operator appearing on the last line as $\mathcal{O}_y\equiv |[O_y]_{\mu\mu'}|$. Notably it is invariant under complex conjugation. Using assumptions (ii)+(iii) and the invariance under complex conjugation we find 
\begin{equation}
\langle \lambda |\mathcal{O}_y | \lambda' \rangle =   2e^{i\varphi}ab \left[\langle \psi_1 | \mathcal{O}_y | \psi_1 \rangle - \langle \psi_2 | \mathcal{O}_y | \psi_2 \rangle\right]  = 0. 
\end{equation}
Grouping the last three equations together we find that 
\begin{align}
0 = \langle \lambda | \mathcal{O}_y | \lambda' \rangle &\geq |\langle \lambda | O_y | \lambda' \rangle| > 0. 
\end{align}
Thus the $K$ symmetry breaking assumptions are incompatible with having two non-negative ground states.  
 
Next, we consider the case when $|\lambda'|<|\lambda|$ where $|\lambda\rangle$ is still non-negative, but now $|\lambda'\rangle$ is only guaranteed to be real. As $|\lambda'|$ and $|\lambda|$ become exponentially close it is natural to expect some notion of continuity in the sense that almost degenerate dominant eigenvectors should also be almost non-negative. If so the previous proof can be repeated. 

Let us establish this almost non-negative behavior for the case where ${\rm T}$ is symmetric and therefore also Hermitian. To this end, split the real state $|\lambda'\rangle$ into two non-negative states with distinct supports -- those with non-negative coefficients $|+\rangle$, and those with negative coefficients $|-\rangle$ such that $|\lambda'\rangle = |+\rangle - |-\rangle$. Note that the states $|+\rangle,|-\rangle$ are not yet assumed to be normalized. Consider next the state $||\lambda'| \rangle = |+\rangle + |-\rangle$ and expand it in the eigenvectors of ${\rm T}$ 
\begin{align}
||\lambda'|\rangle &= |+\rangle + |-\rangle = \sum_{l} c_{l} |l \rangle.
\end{align} 
Due to the non-negative nature of ${\rm T}$, $|| {\rm T} | |\lambda'|\rangle || \geq ||{\rm T} |\lambda'\rangle||$. This, the exponential degeneracy of $|\lambda|$ and $|\lambda|'$, and the $\delta(L)$ gap to other states implies that $\sum_{l \neq \lambda,\lambda'} |c_{l}|^2$ is exponentially small in system size. We thus find 
\begin{align}
\label{Eq:+}
|+\rangle &= \frac{1}{2}\left[(c_{\lambda'}+1) |\lambda' \rangle + c_{\lambda} | \lambda \rangle \right] + O(e^{-L}) \\ 
|-\rangle &= \frac{1}{2}\left[(c_{\lambda'}-1) |\lambda' \rangle + c_{\lambda} | \lambda \rangle \right] + O(e^{-L}) 
\end{align}
Since $|\lambda'\rangle$ is normalized, at least one of these vectors has a norm order of 1, and let us assume without loosing generality that this is the vector $|+\rangle$. Now, depending on the remaining vector, there are two possible scenarios: (s1) $\langle - | - \rangle > e^{-L}$ and (s2) $\langle - | - \rangle < e^{-L}$. In the  case (s1) the joint weights $|c_\lambda'\pm1|^2+|c_\lambda|^2$ are still exponentially larger than the weights of remaining terms. Consequently $|-\rangle$ can be chosen as the second basis vector for the ground-state subspace of $T$. Since $|-\rangle$ and $|+\rangle$ are, by definition, non-negative, and orthogonal one can repeat the previous argument about the expectation value of $\mathcal{O}_y$ with these two states. 

In the  case (s2) we use orthogonality of $|\lambda\rangle$ and $|\lambda'\rangle$ to obtain that $\langle \lambda |+\rangle = \langle \lambda | - \rangle$, and using $\langle \lambda | -\rangle \leq \sqrt{\langle -|-\rangle}$ we finally obtain $\langle \lambda | + \rangle < e^{-L/2}$. We thus take $|\lambda \rangle$ and $|+\rangle$ to be the two dominant vectors. Notably they are both non-negative, $K$ invariant and are, up to exponential correction span the ground state manifold, allowing us to apply the previous argument. 

The essential place we have used ${\rm T} = {\rm T}^{T}$ in the above is when we took the overlap $\langle \lambda | \lambda'\rangle$. For a positive and symmetric ${\rm T}$, $\langle \lambda | = [|\lambda \rangle]^T$ on $\sigma$, however in general it is some dual vector whose amplitudes, while still strictly positive, may not be identical to those of $| \lambda \rangle$. 

To see what changes in the case of non-symmetric $\mathrm{T}$ let us consider a vector $|v\rangle$ with components $v_\sigma= \sqrt{\langle \sigma | \lambda \rangle / \langle \lambda | \sigma \rangle}$, where $\langle \lambda |$ now refers to the dual of $|\lambda \rangle$. The entries of vector $|v\rangle$ are all equal to $1$ in the symmetric case. Consider now a diagonal matrix $\hat{v}$ which diagonal is given by $v_\sigma$. Next perform the similarity transformation 
\begin{align}
{\rm \tilde{T}} = \hat{v} {\rm T} [\hat{v}]^{-1}
\end{align}
If ${\rm T}$ is strictly positive then also is ${\rm T'}$, and clearly they have the same spectrum. What was gained is that the left and right dominant eigenvectors $|\tilde{\lambda}\rangle$, and $\langle \tilde{\lambda}|$ are now given by $\langle \sigma | \tilde{\lambda} \rangle = \sqrt{\langle \sigma | \lambda \rangle \langle \lambda | \sigma \rangle}$ and its transpose. To the extent that $\hat{v}$ can be viewed as a local similarly transformation, the previous proof can now be repeated. However we cannot refute the possibility that $\hat{v}$ is some non-local transformation which can in principal transform a Schrodinger-cat state into a physical state, rendering the previous arguments void. 

\section{Note on the derivation of the gravitational anomaly} 
Here we discuss how to introduce a real twist operator on the single edge. In addition we show that despite of this positivity, the partition function acquires a non-trivial phase factor originating from the gravitational anomaly. 

First, consider the fermionic integer QHE setting defined by the single-particle lattice Hamiltonian $\hat{H}-\mu$ such that $\mu$ is within the band-gap, and all bands below the gap have Chern number $1$. When placed on a cylinder the system will show chiral, and anti-chiral edges states on the opposite edges. Focusing on, say, the right edge, and taking a very long cylinder one can clearly distinguish the states associated with the left and right edge. Let us denote by $|k,r\rangle$ the first-quantized wave-functions of the states on the right edge, and introduce a cut-off operator 
\begin{align}
\hat{C}_{r} &= \sum_k G_{\epsilon,\delta}(k-k_{\mu}) | k, r\rangle \langle k, r |
\end{align} 
where $k_{\mu}$ is the momentum corresponding to the state with energy $\mu$, and $G_{\epsilon,\delta}(x)$ is given by $G_{\epsilon,\delta}(x) = e^{-\epsilon |x| }$ for $|x| < \Lambda$ where $e^{-\epsilon \Lambda } = \delta$,  and zero for larger $|x|$. Next, we write a projected single-particle Hamiltonian on the right edge as 
\begin{align}
\hat{H}_{r} &= \hat{C}^{\dagger}_{r} [\hat{H}-\mu] \hat{C}_{r}
\end{align}
similarly we introduce a projected momentum operator.
\begin{align}
\hat{P}_{r} &= \hat{C}^{\dagger}_{r} \hat{P} \hat{C}_{r}.
\end{align}
The operator $\hat{P}$ is defined by $i a^{-1} \log(T)$ where $T$ is the translation operator along one unit-cell and $a$ is the lattice spacing. Specifically we consider a local basis $|x,\alpha\rangle$ where $x$ denotes a site and $\alpha$ a unit-cell index and construct $\hat{P}$ as 
\begin{align}
\label{Eq:hatP}
\hat{P} &= \sum_{k,\alpha} k | k,\alpha \rangle \langle k,\alpha|  
\end{align}
where $k$ is chosen to be in the first Brillouin zone, and $|k,\alpha \rangle \propto \sum_x e^{i k x} |x,\alpha \rangle$.  

Using projected single particle Hamiltonian, and momentum operator we construct the corresponding many body Hamiltonian ${\rm \hat{H}}_r$, and the many body momentum operator ${\rm \hat{P}}$. Now, let us consider the many-body partition function with a twist in the boundary conditions,
\begin{align}
Z_r[\beta,\phi] &= \mathrm{Tr}[e^{-\beta {\rm \hat{H}_r}}e^{i \phi {\rm \hat{P}_r}}],
\end{align}
where $e^{i \phi {\rm \hat{P}_r}}$ realizes the twist. 

In the limit of very large $\beta$ the partition function will be dominated by the ground state contribution 
\begin{align}
Z_r &= e^{-\beta E_0 + i \phi P_0}.
\end{align}
Taking $k_{\mu} = 0$ for simplicity and assuming a spectrum $\epsilon_k$ we find by summing over occupied states
\begin{align}
P_0 &=  \sum_{n < 0} k_n G_{\epsilon,\delta}(k_n) 
\end{align}
where $k_n = 2\pi n/L$. Now we can evaluate $P_0$ using Euler-Maclaurin formula, to write the sum over momentum as an integral, which yields 
\begin{equation}
P_0 = \frac{L}{2\pi}\int_{-\Lambda}^{0} dk k G_{\epsilon,\delta}(k) + \frac{2\pi}{L}\frac{1}{12}+ O(L^{-1} \delta\log\delta).
\end{equation}

We thus find that $P_0$ contains a diverging energy density which can be understood as the energy density of an infinite edge plus a finite size Casimir-like correction. The contribution from the gravitational anomaly is contained in the second term. To filter out this contribution we can simply compare two systems with sizes $L$ and $2L$, which gives
\begin{align}
P_0(L) - P_0(2L) &= \frac{1}{L}\frac{\pi}{12},
\end{align}
so that
\begin{align}
\label{Eq:RatioL2L}
\arg \left( \frac{Z_r[\beta,\phi;L]}{Z_r[\beta,\phi;2L]}\right) &= \frac{\phi}{L} \frac{\pi}{12}.
\end{align}

Now we consider the above procedure for an operator $\hat{H}$ in the representation, where all the matrix elements are real. So that its eigenvalue subspaces can all be chosen to have real eigenvectors. As a result $\hat{C}_r$ is a real operator. Notice further that $\hat{P}$ is purely imaginary since 
\begin{align}
[\hat{P}]_{x\alpha,y \beta} &= \delta_{\alpha \beta} \sum_k k e^{i k a (x-y)}  
\end{align}
and for every $k$ in the summand there is a $-k$. Consequently $\hat{P}_r = \hat{C}_r \hat{P} \hat{C}_r$ is purely imaginary in the local basis (or equivalently $e^{i \phi P_r}$ is real and positive). Equation \ref{Eq:RatioL2L} thus consists of only real numbers on the l.h.s. which is in contradiction with the r.h.s. 

Next let us turn to the interacting case. To this end we need to express the effective momentum operator on the right edge, ${\rm P_r}$,  without appealing to a single particle basis. We suggest the following obvious solution, we consider a cylinder of large circumference $L$ such that a rotation in a finite angle $\phi_0$ is equivalent to a translation by many ($L \phi/2 \pi$) unit cells. We then define $e^{i \phi_0 {\rm P_r}}$ to be the operator which rotates the system by $\phi_0$ up to, say, 10 correlations length away from the right edge and then gradually rotates by less and less until at a distance of, say, 20 correlation lengths away from the right edge it stops rotating the reminder of the cylinder. We refer to this as twisted-rotation. Notably the twisted-rotation operator thus defined is strictly positive. 

The quantity of interest is then $\langle gs |e^{i \phi_0 {\rm P_r}}| gs \rangle$ which should contain the same $e^{i \phi \pi/12}$ Casimir term as before. Since it rotates the right edge and not the left edge we expect the edge contribution to yield the same contribution as before. One may worry though that an extra bulk contribution would appear which would cancel the right edge's contribution. We find this unlikely since Casimir effects are $1/L$ effects and these should not occur in a gapped system such as the bulk of a FQHE. This can be verified explicitly using a continuum version of $e^{i \phi_0 {\rm P_r}}$ for the IQHE on a cylinder in the Landau gauge. It is then easy to show that $e^{i \phi_0 {\rm P_r}}$, defined using twisted-rotation, has essentially the same algebraic properties as previous one sided rotation, defined using spectral projections. 

\end{document}